\documentclass[aps,preprint,nofootinbib,tightenlines,12pt]{revtex4}%
\usepackage[french,english]{babel}
\usepackage[paperwidth=210mm,paperheight=297mm,centering,hmargin=2.5cm,vmargin=2.5cm]{geometry}
\usepackage{amsfonts}
\usepackage{amsmath}
\usepackage{amssymb}
\usepackage{graphicx}%
\usepackage{color}
\usepackage[T1]{fontenc}
\usepackage{lmodern}
\usepackage{csquotes}
\providecommand{\U}[1]{\protect\rule{.1in}{.1in}}
\begin{document}

\preprint{ }
\title[ ]{On the Unconditional Validity of J. von Neumann's Proof of the Impossibility of Hidden Variables in Quantum Mechanics}
\author{C. S. Unnikrishnan}
\affiliation{Tata Institute of Fundamental Research, Homi Bhabha Road, Mumbai 400005}

\begin{abstract} 
The impossibility of  theories with  hidden variables as an alternative and replacement for quantum mechanics was discussed by J. von Neumann in 1932. His proof was criticized as being logically circular, by Grete Hermann soon after, and as fundamentally flawed, by John Bell in 1964. Bell's severe criticism of Neumann's proof and the explicit (counter) example of a hidden variable model for the measurement of a quantum spin are considered by most researchers, though not all, as the definitive demonstration that Neumann's proof is inadequate. Despite being an argument of mathematical physics, an ambiguity of decision remains to this day. I show that Neumann's assumption of the linear additivity of the expectation values, even for incompatible (noncommuting) observables, is a necessary constraint related to the nature of observable physical variables and to the conservation laws. Therefore, any theory should necessarily obey it to qualify as a physically valid theory. Then, obviously, the hidden variable theories with dispersion-free ensembles that violate this assumption are ruled out. I show that it is Bell's counter-example that is fundamentally flawed, being inconsistent with the factual mechanics. Further, it is shown that the local hidden variable theories, for which the Bell's inequalities were derived, are grossly incompatible with the fundamental conservation laws. I identify the intrinsic uncertainty in the action as the reason for the irreducible dispersion, which implies that there are no dispersion-free ensembles at any scale of mechanics. With the unconditional validity of its central assumption shown, Neumann's proof is fully resurrected.
\vspace {0.6 in}

\noindent\normalsize Note: A French translation of the title, abstract and a slightly condensed version of the introductory section with a summary of the main results are included in the manuscript (pp 3-6), between the table of contents and the main text in English.

\end{abstract}

\date{\today}
\startpage{1}
\endpage{102}
\maketitle

\tableofcontents

\newpage
\selectlanguage{french}
\begin{center}
		{\large \textbf{Sur la validité inconditionnelle de la preuve de J. von Neumann de l'impossibilité de mécanique quantique à variables cachées}}
		\vspace {0.2 in}
		
		C. S. Unnikrishnan
		
		Tata Institute of Fundamental Research, Homi Bhabha Road, Mumbai 400005
		\vspace {0.1 in}
	\end{center}
\section*{R\'esum\'e}
\vspace{0.1 in}

{L'impossibilité de théories à variables cachées (variables supplémentaires) comme alternative et remplacement de la mécanique quantique fut débattue par J. von Neumann en 1932. Sa preuve fut critiquée peu après par Grete Hermann, qui la qualifia de raisonnement circulaire, puis par John Bell qui la déclara fondamentalement défectueuse en 1964. La critique sévère de Bell contre la preuve de Neumann, et le contre-exemple explicite d'un modèle à variables cachées pour la mesure d'un spin quantique, sont considérés par la plupart des chercheurs, mais non la totalité, comme la démonstration que la preuve de Neumann est inadéquate. Bien que ce soit un argument en physique mathématique, une ambiguïté de décision persiste à ce jour. Je démontre que l'hypothèse de Neumann sur la linéarité additive des valeurs moyennes, même pour des observables incompatibles (qui ne commutent pas), est une contrainte nécessaire relative à la nature des variables physiques observables et aux lois de conservation. Par conséquent, toute théorie devrait obligatoirement lui obéir pour être qualifiée de théorie physiquement valable. Ainsi, de toute évidence, les théories à variables cachées avec des ensembles de dispersion-libre (dispersion-nulle) qui violent cette hypothèse sont éliminées. Je démontre que c'est le contre-exemple de Bell qui est fondamentalement défectueux, étant incohérent avec la mécanique factuelle. De plus, il est démontré que les théories locales à variables cachées, pour lesquelles Bell déduisit les in\'egalités, sont incompatibles de façon flagrante avec les lois fondamentales de conservation. J'identifie l'incertitude intrinsèque dans l'action comme la raison de la dispersion irr\'eductible, ce qui implique qu'il n'y a pas d'ensembles de dispersion-libre à aucun niveau de la mécanique. Avec la validité inconditionnelle de son hypothèse centrale démontrée, la preuve de Neumann est totalement ressuscitée.}
\newpage
\section*{Introduction des principaux résultats}

	\noindent<<\textit{Une seule exigence de (Leonard) Nelson reçut toute mon approbation. C'était l'exigence de ne pas s'empêcher de répondre par peur de se couvrir de honte}.>>\ \ -- Grete Hermann.
\vspace {0.2 in}

Le sujet principal de cet article est la preuve que l'hypothèse centrale de l'additivité linéaire des valeurs moyennes (valeurs attendues), dans la preuve de J. von Neumann \cite{JvNP} sur l'impossibilité de toute description de mécanique quantique à variables cachées (variables supplémentaires), est inconditionnellement valable parce qu'il existe une contrainte physique cruciale. Cet article restitue la validité générale de la preuve de Neumann, qui fut mise en doute par de nombreux chercheurs depuis sa présentation en 1932. Plusieurs autres résultats essentiellement liés au sujet principal sont aussi systématiquement examinés. Le résultat principal est la preuve que l'hypothèse de Neumann sur l'additivité linéaire des valeurs moyennes, $Exp(aR + bS) = aExp(R) + bExp(S)$, est effectivement une relation à laquelle tous les ensembles physiques obéissent, et qui s'applique dans toutes les théories physiques valables, indépendamment du fait que $R$ et $S$ sont, ou non, des quantités physiques mesurables simultanément, ou des observables qui commutent dans la terminologie de la mécanique quantique. Ceci ressuscite la preuve de Neumann de l'impossibilité de toute description de la mécanique quantique à variables cachées. Les failles des diverses critiques de la preuve de Neumann par Grete Hermann et John Bell, sont ensuite examinées en détail. 

Le deuxième résultat est l'échec du célèbre contre-exemple de Bell du théorème de Neumann, un modèle à variables cachées des mesures de la projection spin d'une particule spin-1/2. Ceci est démontré de deux façons différentes, exposant clairement la nature dénuée de signification physique et l'inconsistance de ce contre-exemple. 

Le troisième résultat est une preuve clarifiant que  \guilsinglleft{l'incomplétude de causalité}\guilsinglright, qui motiva le travail sur les théories à variables cachées, est très différente de \guilsinglleft{l'incomplétude selon le raisonnement EPR dans la mécanique quantique}\guilsinglright\ débattue par Einstein, Podolsky et Rosen, en 1935. Une analyse claire de cette distinction était nécessaire car la preuve de Neumann concerne l'incomplétude causale qui est enracinée dans l'indéterminisme, alors que la discussion EPR se rapporte à un aspect entièrement différent de la mécanique quantique qui ne concerne pas l'indéterminisme et les variables cachées, bien qu'il y ait communément confusion avec la première notion.  

Le quatrième résultat identifie que la relation entre la dispersion quantique et la fonction d'action est la raison exacte de l'impossibilité d'ensembles à dispersion-libre (dispersion-nulle). Ce résultat anticipe toute idée fausse qu'une description déterministe de la mécanique serait un jour possible à un quelconque degré, exceptée en tant qu'approximation.

Le cinquième résultat est que les théories locales à variables cachées, pour lesquelles Bell déduisit les inégalités, sont, dans les faits, incompatibles avec les lois fondamentales de conservation, et par conséquent ce sont des théories non physiques dans leur formulation même.  

Une brève analyse de la réinterprétation que fit D. Bohm \cite{Bohm-universe} de la mécanique quantique de Schrödinger en tant que théorie non locale {\guilsinglleft à variables cachées\guilsinglright} est inclue pour que cet article soit complet, montrant certaines de ses sérieuses insuffisances.

En tant que résultat de la physique mathématique, on s'attend à ce que la preuve de Neumann soit par déduction une suite des hypothèses utilisées dans la preuve. 
Par conséquent, le seul moyen de mettre la preuve en doute est de questionner la validité des hypothèses de départ. C'est en fait ce que firent à la fois Grete Hermann en 1933-35 \cite{Hermann-history,Grete35} et John Bell en 1964-66 \cite{Bell-RMP}, quand ils critiquèrent l'hypothèse centrale de Neumann, selon laquelle la fonction de valeur moyenne $Exp(O)$ obéit à la linéarité additive dans toutes les théories, même pour les observables de type $O = aR+bS$, où $R$ et $S$ ne commutent pas en mécanique quantique. La fonction de {\guilsinglleft valeur moyenne\guilsinglright} (valeur attendue) est $Exp(O)=\langle \psi|\hat{O}|\psi\rangle $ en mécanique quantique, où $\hat{O}$ est l'opérateur hermitique correspondant à l'observable $O$. Dans une théorie générale, et aussi de manière empirique, $Exp(O)$ est une moyenne (valeur moyenne) des valeurs possibles de la quantité $O$ dans un ensemble statistique. Chaque quantité physique $O$ a une valeur moyenne $Exp(O)$ et une variance $Var(O)=Exp(O-\bar{O})^{2}=Exp(O^{2})-\bar{O}^{2}$. 
Un ensemble pour lequel $Var (O) = 0$  pour toutes les observables est défini comme un ensemble de {\guilsinglleft dispersion-libre\guilsinglright} \cite{JvNP}. 

La relation de linéarité additive 
\begin{equation}\label{key}
	Exp(aR + bS) = aExp(R) + bExp(S) 
\end{equation} 
est bien entendu valable pour des ensembles classiques, où la quantité $(aR + bS)$ se compose des  mesures séparées de $R$ et $S$. Cela est également valable en mécanique quantique, même pour les observables qui ne commutent pour lesquelles $(aR+bS)$ n'est défini qu'implicitement. C'est à dire qu'une valeur pour $O = (aR+bS)$ ne peut pas se composer des mesures de $R$ et $S$ si les quantités ne sont pas simultanément mesurables. Pourtant, on qualifie $(aR+bS)$ ainsi que la quantité plus générale $(af(R)+bg(S))$  {\guilsinglleft d'observables\guilsinglright} en mécanique quantique.  
Un exemple pratique est $L_n$, la composante du momentum angulaire (à deux dimensions) dans la direction $\hat{n}=a\hat{x}+b\hat{y}$. En mécanique quantique, l'opérateur pour cette composante est $\hat{L}_{n}=a\hat{L}_{x}+b\hat{L}_{y}$, mais la valeur de $L_n$ ne peut pas être obtenue des mesures de $L_x$ et $L_y$. 

Pourtant, les valeurs moyennes des trois quantités obéissent au principe de linéarité additive. En fait, Neumann prit soin de prouver l'additivité linéaire en mécanique quantique de manière explicite, car sa validité n'était pas évidente \cite{JvNP}. Comme les opérateurs hermitiques de la mécanique quantique obéissent au principe de linéarité additive, 
\begin{equation}
	Exp(O)\equiv Exp(aR+bS)=\langle \psi|a\hat{R}+b\hat{S}|\psi\rangle
	=\langle \psi|a\hat{R}|\psi\rangle +\langle \psi|b\hat{S}%
	|\psi\rangle =aExp(R)+bExp(S)
\end{equation}
Ici, l'opérateur $a\hat{R}+b\hat{S}$ représente une seule observable (quantité physique) comme $\hat{L}_{n}$, devant être mesurée par un dispositif adapté, qui en général est différent de ce qui est approprié pour mesurer $R$ ou $S$. Le reproche formulé dans les critiques était que Neumann n'avait pas raison d'assumer que la validité de la linéarité additive s'appliquait aussi dans les théories à variables cachées, car de telles théories implémentent la distribution de résultats physiques sous des prémisses très différentes, en comparaison avec la mécanique quantique. Selon les termes de Bell \cite{Bell-RMP},
\begin{quote}
	...L'additivité des valeurs attendues... est une propriété très particulière des états de la mécanique quantique, que l'on n'attend pas a priori. Il n'y a aucune raison de l'exiger individuellement des états hypothétiques de dispersion-libre, dont la fonction est de reproduire les particularités mesurables de la mécanique quantique quand la moyenne est faite.
\end{quote} 

C'est le point central sur lequel je me focalise et je vais prouver que l'additivité des valeurs moyennes est effectivement une propriété essentielle qui est exigée a priori, et qui est universellement nécessaire, dans toutes les théories physiques. Etant donné qu'un ensemble de dispersion-libre ne peut pas répondre à cette contrainte, de tels ensembles n'existent pas dans le monde physique. Ceci exclut définitivement toute description de la mécanique quantique à variables cachées et justifie pleinement la preuve de Neumann.

Les théories à variables cachées complètent l'état pur en mécanique quantique (la fonction-$\psi$) avec des variables cachées qui déterminent une relation causale entre les valeurs des variables et la valeur d'une quantité physique observée dans les faits. 
Le but est d'obtenir les valeurs quantiques aléatoires, apparemment acausales, comme le résultat déterministe de l'état quantique et des valeurs stochastiques des variables cachées. 
Bien sûr, il est empiriquement connu et vérifié que la relation d'additivité est valable lorsque les résultats de l'ensemble complet des mesures sont pris en considération pour calculer la moyenne. Cependant, a-t-on raison de supposer sa validité également pour chaque sous-ensemble?  Cette question se pose car la relation fonctionnelle allant des valeurs des variables cachées à la valeur observée d'une quantité physique pourrait être non-linéaire.
Alors que G. Hermann jugeait la preuve de Neumann d'être <<logiquement circulaire>>, J. S. Bell alla jusqu'à qualifier la preuve (dont l'abréviation est JvNP) de <<idiote>>, <<absurde>> et <<folle>> \cite{Bell-pilot,Dieks}. Pour défendre son point de vue, Bell mit au point un contre-exemple à la JvNP, un modèle à variables cachées des mesures d'un seul spin quantisé avec comme résultat la valeur moyenne correcte \cite{Bell-RMP,Bell64}. Comme on le sait bien, il alla plus loin pour démontrer qu'un tel modèle n'était pas possible pour la corrélation générale de deux spins, dans l'hypothèse de la localité. 

Aujourd'hui, dans la communauté de physiciens la plupart se fient à la critique sévère de Bell, essentiellement parce qu'il fit une démonstration explicite contre la preuve de Neumann, d'un modèle à variables cachées des mesures d'un seul spin. Pourtant, il y eut de sérieux débats et désaccords \cite{Dieks,Bub,Mermin}. 
Les énormes efforts et ressources qui furent investis pour tester les théories locales à variables cachées (LHVT) et les inégalités de Bell provenaient du fait que Bell r\'efuta la JvNP, et par ailleurs du résultat qu'il n'y a aucune différence distincte entre les fonctions de corrélation à deux particules et la mécanique quantique. D'un autre côté, si la JvNP avait toujours été valable et que Bell se soit trompé dans ses critiques et contre-exemple, nous serions alors face à une situation de recherche et tests expérimentaux malavisés depuis des décennies. 

Nous allons voir que la linéarité additive de la fonction $Exp(O)$ découle de la nécessité que les observables physiques obéissent aux lois fondamentales de conservation dans tout ensemble. Il s'ensuit alors la preuve qu'il n'y a pas d'ensembles de dispersion-libre. Ainsi, la partie manquante de la JvNP est en place, découlant de considérations de physique fondamentale. 
Par conséquent, la critique de raisonnement circulaire de Hermann et la critique d'absurdité de Bell ne sont pas valables. Avec la JvNP démontrée comme généralement valable, il est possible d'identifier le point exact où Bell se trompe dans sa critique. De plus, je vais montrer explicitement que le contre-exemple de Bell du modèle à variables cachées de la mesure du spin  est fondamentalement défectueux, et qu'il est incohérent avec la mécanique quantique simple du spin. 
Dans toutes les analyses précédentes du problème, la raison principale de l'impossibilité de tout ensemble à dispersion-libre ne fut pas explorée. Je fais remonter cette raison à la relation profonde entre la fonction d'action d'Hamilton et la dispersion irréductible des variables dynamiques en mécanique.

Je montre que la dispersion irréductible n'est pas dans les observables individuelles comme $p$, $q$, $L$, etc., mais dans la fonction d'action même, qui est une combinaison de deux observables conjuguées (incompatibles). Comme l'action $\mathcal{S}$ est la même pour n'importe quel sous-ensemble d'un ensemble parent défini par une fonction-$\psi$, la preuve de  l'impossibilité de théories à variables cachées avec des ensembles à dispersion-libre est immédiate.

\newpage
\selectlanguage{english}
\section{Introduction to the Main Results}

\noindent``\emph{Only one demand that (Leonard) Nelson made got
my total approval. It was the demand not to let oneself be prevented from
answering for fear of disgrace}''\ \ -- Grete Hermann.

\medskip

The main theme of this paper is the proof that the central assumption of the
linear additivity of expectation values, in J. von Neumann's proof \cite{JvNP}
of the impossibility of any hidden variable description of quantum mechanics,
is unconditionally valid because of a crucial physical constraint. This
restores the general validity of Neumann's proof that has been questioned by
many researchers after its presentation in 1932. Several supplementary results
that are vitally related to the main theme are also discussed systematically.
The central result is the proof that Neumann's assumption of the linear
additivity of the expectation values, $Exp(aR+bS)=aExp(R)+bExp(S)$, is indeed
a relation obeyed by all physical ensembles, and applicable in all physically
valid theories, irrespective of whether or not $R$ and $S$ are physical
quantities that are simultaneously measurable, or commuting observables in the
terminology of quantum mechanics. This resurrects Neumann's proof of the
impossibility of any hidden variable description of quantum mechanics (QM).
The fault lines of the different criticisms of Neumann's proof, by Grete
Hermann and John Bell, are then shown in detail. 

The second result is the
failure of Bell's well known counter-example to Neumann's theorem, a hidden
variable model of the measurements of the spin projection of a spin-1/2
particle. This is demonstrated in two different ways, exposing clearly the
unphysical nature and inconsistency of that counter-example. 

The third result
is a clarificatory proof that the `causality incompleteness' that motivated
the work on hidden variable theories is very different from the
`EPR-incompleteness' of QM, discussed by Einstein, Podolsky, and Rosen, in 1935. This
was necessary because Neumann's proof concerns the `causal incompleteness' that
is rooted in indeterminism, whereas the EPR discussion is on an entirely
different aspect of QM that does not concern indeterminism and hidden
variables, though commonly confused with the former notion. 

The fourth result
identifies the exact reason for the impossibility of dispersion-free
ensembles, in the relation between the quantum dispersion and the action
function. This result pre-empts any misconception that a deterministic
description of mechanics will ever be possible at any scale, except as an
approximation. 

The fifth result is that the local hidden variable theories for
which the Bell's inequalities were derived are factually incompatible with the
fundamental conservation laws, and hence they are unphysical theories in their
very formulation.

A brief discussion of D. Bohm's reinterpreation of the Schr\"odinger quantum mechanics as a nonlocal theory \cite{Bohm-universe}, with the position coordinate as a `hidden variable', is included for completeness, pointing out some of its serious inadequacies.

Seen as a result of mathematical physics, one expects that Neumann's proof
deductively follows from the assumptions used in the proof. Therefore, the
only way the proof could be put in doubt is by questioning the validity of the
starting assumptions. In fact, this is what was done by both Grete Hermann in
1933-35 \cite{Hermann-history,Grete35} and John Bell in 1964-66
\cite{Bell-RMP}, when they criticized Neumann's central assumption that the
expectation value function, $Exp(O)$, obeys linear additivity in all theories,
even for the observables of the form $O=aR+bS$, where $R$ and $S$ are
noncommuting in quantum mechanics. The function `expectation value' is
$Exp(O)=\langle \psi|\hat{O}|\psi\rangle $ in quantum mechanics,
where $\hat{O}$ is the Hermitian operator corresponding to the observable $O$.
In a general theory, and also empirically, $Exp(O)$ is an average (mean value)
of the possible values of the quantity $O$ over a statistical ensemble. Every
physical quantity $O$ has an expectation value $Exp(O)$ and a variance
$Var(O)=Exp(O-\bar{O})^{2}=Exp(O^{2})-\bar{O}^{2}$. \emph{An ensemble for
which }$Var(O)=0$\emph{ for all observables is defined as a `dispersion-free'
ensemble }\cite{JvNP}.

The linear additivity relation
\begin{equation}
Exp(aR+bS)=aExp(R)+bExp(S)
\end{equation}
is of course valid for classical ensembles, where \emph{the quantity
}$(aR+bS)$\emph{ is assembled from separate measurements of }$R$\emph{ and
}$S$\emph{. It is also valid in quantum mechanics (QM), even for noncommuting
observables for which }$(aR+bS)$\emph{ is defined only implicitly. }That is, a
value for $O\equiv(aR+bS)$ cannot be formed from the measurements of $R$ and
$S$ if the quantities are not simultaneously measurable. Yet, $(aR+bS)$ as
well as the more general $(af(R)+bg(S))$ qualify as `observables' in quantum
mechanics. One example that Neumann himself discussed is the energy
observable, $E=\left(  p^{2}/2m\right)  +V(x)$. Another common and convenient
example is $L_{n}$, the component of the (two dimensional) angular momentum
along the direction $\hat{n}=a\hat{x}+b\hat{y}$. In QM, the operator for this
component is $\hat{L}_{n}=a\hat{L}_{x}+b\hat{L}_{y}$, but the value of $L_{n}$
cannot be obtained from the measurements of $L_{x}$ and $L_{y}$. Yet, the
expectation values of the three quantities obey the principle of linear additivity. In
fact, Neumann took care to prove the linear additivity in quantum mechanics
explicitly, because its validity was not obvious \cite{JvNP}. The proof within
the mathematical structure of QM is easy because the Hermitian operators of QM
obey the principle of linear additivity,
\begin{equation}
Exp(O)\equiv Exp(aR+bS)=\langle \psi|a\hat{R}+b\hat{S}|\psi\rangle
=\langle \psi|a\hat{R}|\psi\rangle +\langle \psi|b\hat{S}%
|\psi\rangle =aExp(R)+bExp(S)
\end{equation}
Here, the operator $a\hat{R}+b\hat{S}$ represents a single observable
(physical quantity) like $\hat{L}_{n}$, to be measured by a suitable apparatus
that is in general different from what is appropriate for measuring $R$ or $S$.

The complaint of the critics was that Neumann was not justified in assuming
the validity of the linear additivity in the hidden variable theories as well,
because such theories implement the distribution of observable physical
results under very different physical premises, compared to QM. In Bell's
words \cite{Bell-RMP},

\begin{quote}
...the additivity of expectation values.... is a quite peculiar property of
quantum mechanical states, not to be expected a priori. There is no reason to
demand it individually of the hypothetical dispersion-free states, whose
function is to reproduce the \emph{measurable} peculiarities of quantum
mechanics \emph{when averaged over}.
\end{quote}

This is the central point that I focus on, and I will prove that \emph{the
additivity of the expectation values is indeed an essential property that is
demanded a priori, and universally necessary, in all physical theories}. Since
a dispersion-free ensemble cannot fulfill this constraint, such ensembles do
not exist in the physical world. This definitely rules out any hidden variable
description of quantum mechanics and fully justifies Neumann's proof.

The hidden variable theories supplement the pure quantum mechanical state (the
$\psi$-function) with a set of hidden variables that determine a causal
relation between the values of the variables and the actually observed value
of a physical quantity. The aim is to get the random, apparently acausal,
quantum values as a deterministic outcome from the quantum state and the
hidden stochastic values of the hidden variables. Of course, it is empirically
known and verified that the additivity relation is valid when the entire
ensemble of measurement results are considered to calculate the average.
However, is one justified in assuming its validity for every subensemble as
well? This question arises because the functional connection from the values
of the hidden variables to the observed value could be nonlinear. While G.
Hermann judged Neumann's proof as \textquotedblleft logically
circular\textquotedblright, J. S. Bell went to the extent of calling the proof
(abbreviated as JvNP) as ``silly'',
``absurd'', and ``foolish'' \cite{Bell-pilot,Dieks}. To prove his point, Bell
constructed a counter-example to the JvNP, a hidden variable model of the
measurements of a single quantized spin that resulted in the correct
expectation value \cite{Bell-RMP,Bell64}.  As it is well known, he went
further to show that such a model was not possible for the general correlation
of two spins, under the assumption of locality. Recent discussions that review
the situation differ in their judgement of the JvNP and Bell's criticism. More
balanced views describe the JvNP as of limited validity, applicable to only a
certain class of hidden variable theories \cite{Bub}.

It is curious to note that E. P. Wigner's account \cite{Wigner-AJP} of the
hidden variable question judged that \emph{Bell had arrived at the same
conclusion as Neumann's}, though with the aid of a different argument! This
statement is correct in a sense, as far as the final results are concerned.
The article, written in 1969, seems to fully accept Neumann's reasoning, as
well as Bell's, and does not express any doubt about the integrity of
Neumann's proof. However, Wigner mentions that Neumann was convinced against
hidden variables primarily because of physical reasons associated with
multiple (sequential) quantum measurements. This is an important point,
because the repeated measurements of a pair of incompatible observables do
show the difficulty in proposing the existence of ensembles without dispersion
(variance) in the physical quantities. However, in this paper we are concerned
with Neumann's formal proof and the major criticisms that were mentioned.

In the history of the discussions of the JvNP, there is another curious fact,
that Hermann's criticism did not reach Neumann, even though she sought
comments from researchers like Bohr, Heisenberg and Dirac
\cite{Hermann-history}. Bell's criticisms were expressed after Neumann's
departure. Therefore, we do not know what justification Neumann would have
given for taking the linear additivity of the function $Exp(O)$ as generally
applicable for all statistical ensembles. Though Neumann discusses the issues
in great detail in his monograph, he is terse at times, and seems to avoid
explanatory comments on the mathematical steps that are perhaps evident to
him. In any case, instead of trying to interpret Neumann's reasons, I will
present a strengthened proof, without the alleged ambiguities. At present,
most people in the physics community trust Bell's harsh criticism,
\emph{primarily because of his explicit demonstration countering Neumann's
proof}, of a hidden variable model of the measurement on a single spin.
However, there has been some serious debates and disagreements, involving J.
Bub, D. Dieks and D. Mermin, regarding Hermann's and Bell's criticisms about
Neumann's proof \cite{Dieks,Bub,Mermin}. The enormous efforts and resources
that had been invested in the tests of local hidden variable theories (LHVT)
and the Bell's inequalities were due to Bell's refutation of the JvNP, along
with the result that there was a distinct difference between the measurable
two-particle correlation functions in the LHVT and QM. On the other hand, if
the JvNP was always valid, and Bell was wrong in his criticisms and
counter-example, then we are faced with a situation of decades' long misguided
research and experimental tests.

I will show that the linear additivity of the function $Exp(O)$ holds for all
physical observables, in any physically valid theory, because every physical
observable should be consistent with the space-time symmetries and should
\emph{obey the fundamental conservation laws on the average}. Then the proof
that there are no dispersion-free ensembles follows. Thus, the missing part in
the JvNP is in place, derived from fundamental physical considerations.
Consequently, Hermann's criticism of logical circularity and Bell's criticism
of absurdity are not valid. With the JvNP shown as generally
valid, one is able to show where exactly Bell's criticism is mistaken.
Further, I will explicitly show that Bell's counter-example of the hidden
variable model of spin measurements was fundamentally flawed, and that it is
inconsistent with the simple quantum mechanics of the spin.

In all previous discussions of the problem, the core reason for the
impossibility of any dispersion-free ensemble was not explored. However, if
one can prove such an impossibility from a general consideration obeyed by all
valid theories, then there should be a deep and fundamental reason that could
be stated simply. I trace this reason to the relation between Hamilton's
action function and the irreducible dispersion of dynamical variables in
mechanics. I show that the irreducible dispersion is not in the individual
observables like $p$, $q$, $L$ etc., but in the action function itself, which is a
combination of two conjugate (incompatible) observables. Since the action $\mathcal{S}$
is the same for any of the subensembles of a parent ensemble defined by a
$\psi$-function, the proof of the impossibility of 
hidden variable theories with dispersion-free ensembles is immediate.

\section{The One Crucial Difference}

I want to state very clearly the one crucial difference between quantum
mechanics and a hidden variable theory, both of which rely on the $\psi
$-function for the representation of a physial state. In QM, the $\psi
$-function is the sole representation of the physical state, from which all
observable results follow. The result of any single observation is not
predictable even though the possible values and their statistical distribution
are calculated. In contrast, in a hypothetical hidden variable theory, the
$\psi$-function as well as many hidden variables together \emph{determine
causally} the precise observable values of the physical quantities, in each
observation. The unpredictability arises merely from the unknown values taken by
the hidden variables. The implication is that the physical nature is
fundamentally deterministic and completely causal, and that at least a
theoretical picture of that hypothetical deterministic world can be
constructed, in a hidden variable theory. Stated this way, the drastic nature
of Neumann's mathematical assertion, that this idealistic expectation of the
physical world is impossible, is striking.

The physical state in QM is \emph{completely} specified by the $\psi
$-function, in the sense that there is a one-to-one correspondence between the
physical states and their $\psi$-function representations. But, the same state
can be expressed in different basis states. When the two sets of basis states
correspond to (eigenstates of) noncommuting observables, then each eigenstate
in one basis is a superposition of eigenstates of the other basis. Denoting
the two basis sets as $\psi$ and $\phi$, a pure state $\psi_{i}$ is in general
\
\begin{equation}
\psi_{i}=\sum\limits_{j}c_{j}\phi_{j} \label{superpose}%
\end{equation}
\emph{This defines the dispersion in the }$\psi$\emph{-state, of the values
represented by the }$\phi_{j}$\emph{ states}. Clearly, every particle in the
$\psi_{i}$ state belongs to the single `homogeneous' ensemble; $\psi$ is not a
`mixture' of ensembles of different $\phi_{j}$ states. Similarly, each
$\phi_{i}$ state is a linear superposition of the different $\psi_{j}$
states. The ensemble specified by the $\psi$-function (or the $\phi$-function)
is dispersive in general; $Var(O)\neq0$ for some observables and hence, there
are no dispersion-free ensembles in quantum mechanics.

In a hidden variable theory, the dispersion of a physical quantity in the
$\psi$-ensemble is seen as due to the mixing of several pure (dispersion-free)
$\phi$-ensembles. Similarly, the dispersion in a pure $\phi$-ensemble results
from a mixing of different $\psi$-ensembles. Hence, the hidden variable
theories assert the notional division of the entire ensemble into
dispersion-free ensembles, $\left(  \psi_{i},\phi_{j}\right)  $, each of which
has no dispersion in any of the observables. This also means that the
$\psi_{i}$ state is not a superposition of the $\phi_{j}$ states. (I use the
term `superposition' exclusively for expressions of the kind in equation
\ref{superpose} and the term `mixture' to refer to the composition of
ensembles. Neumann has used the word `superposition' for both, with its
contextual meaning, which I avoid). This absence of superposition in a hidden
variable theory, and hence the absence of the concept of wave-particle duality
that defines quantum mechanics, is the crucial conceptual difference to be
noted. A hidden variable theory tries to match the measured dispersion in the
given state $\psi$ to what is predicted by quantum mechanics, by distributing
the measured values among the different dispersion-free sets. Each
measurement results in a specific value among the many possible values
randomly because of the hidden variables taking a random (unknown)
realization, which then picks one among the many dispersion-free ensembles. I
want to emphasize the point that \emph{the characteristic feature of a hidden
variable theory is the existence of dispersion-free ensembles}, and not merely
the labelling of some unmeasured physical observables as hidden variables.

Both in QM and in any hidden variable extension, the following is true: A
single measurement of a physical quantity $O$ in a general state specified by
a $\psi$-function returns one of the possible values $a_{i}$ in the spectrum
of that quantity, realized unpredictably. In QM this is one of the eigenvalues
of the operator $\hat{O}$ corresponding to the observable $O$. Then the
expectation value in $n$ measurements is $Exp(O)=\frac{1}{n}\sum a_{i}$. If
the physical state is specified by a $\psi$-function that is an eigenstate
$\psi_{i}$ of an observable $O$, with the eigenvalue $\alpha$, then the
measurable as well as the measured values of $O$ are all $\alpha$, with
$Exp(O)=\alpha$, irrespective of the values of the hidden variables. This will
be important when we discuss Neumann's proof as well as Bell's counter-example.

Another relevant fact that one should keep in mind is that Neumann was asking
the question \emph{whether quantum mechanics could be supplemented by
additional variables} to make it causal. Hence, he assumed that all the
mathematical truths and the statistical predictions from the Schr\"{o}dinger
equation within the theory of quantum mechanics remain true. For example, the
statement that every observable can be represented by a Hermitian operator
must remain true in a hidden variable extension of QM. Then it is also true
that if $\hat{R}$ and $\hat{S}$ are Hermitian operators, so is their sum $\hat{R}+\hat{S}$. However,
we accept the general relation $Exp(aR+bS)=aExp(R)+bExp(S)$ only for the full
ensemble a priori (which agrees with the empirical fact), and not for the
hypothetical subensembles which might be dispersion-free. It might be relevant
to state that the `causal extension' by supplementing QM with hidden variables
is very different from a `total replacement' by a classical hidden variable
theory, without a $\psi$-function description of the physical state.

\section{John von Neumann's Impossibility Proof}

Neumann's proof of the impossibility of a hidden variable description of
quantum mechanics has been discussed extensively. The question addressed by
Neumann was whether a causally deterministic description of quantum mechanics
was possible, by adding hidden supplementary variables to the QM description
of the state specified by the $\psi$-function. In a causal and fully
deterministic mechanics, one can form subensembles in which the values of the
physical quantities are without any variance. In such an ensemble, the
observables have one of their characteristic quantized values or eigenvalues
(in the vocabulary of quantum mechanics). So, the possibility of
dispersion-free ensembles characterize the hidden variable theories. 

Neumann starts with the central assumption, provable as valid in quantum
mechanics, of the general (unconditional) linear additivity of expectation
values of the physical quantities $R,S,$ etc.,
\begin{equation}
Exp(aR+bS)=aExp(R)+bExp(S)
\end{equation}
Also, it is assumed that if the observable values of $O$ are non-negative,
then $Exp(O)\geq0$. Then he argued that there are no dispersion-free ensembles
\cite{JvNP}. First he showed that the function $Exp(O)$ can be expressed in
general as the trace of the product of a Hermitian density matrix $D$ and
another Hermitian matrix $O$ representing the observable, $Exp(O)=Tr(DO)$. A
valid density matrix obeys $Tr(D)=1$.

By the assumption of linear additivity
\begin{equation}
Exp(aR+bS)=aTr(DR)+bTr(DS)
\end{equation}
For a dispersion-free ensemble $\left[  Tr(DO)\right]  ^{2}=Tr(DO^{2})$. Now,
take $O=P_{\phi}$, which is the projection operator onto the pure state $\phi
$. Since $P_{\phi}=P_{\phi}^{2}$, $Tr(DP_{\phi})=\left\langle \phi
|D|\phi\right\rangle $ in the state $\phi$, and $\left\langle \phi
|D|\phi\right\rangle =\left\langle \phi|D|\phi\right\rangle ^{2}$, because the
ensemble is dispersion-free. This implies that $\left\langle \phi
|D|\phi\right\rangle =1$ or $0$. Since this should vary continuously when one
goes continuously from a normalized state $\phi$ to another $\phi^{\prime}$,
$\left\langle \phi|D|\phi\right\rangle $ is a constant. The value
$\left\langle \phi|D|\phi\right\rangle =0$ for all states $\phi$ is physically
irrelevant. Then we conclude that the matrix $D$ is the unit matrix, $D=1$,
for a dispersion-free ensemble. But this is not a valid (probability) density
matrix. Hence, dispersion-free ensembles do not exist and the hidden
variable theories are impossible.

It is clear that \emph{if Neumann's impossibility proof is generally valid,
the entire exercise of discussing and testing hidden variable theories during
the past half century has been a colossal misadventure}. Therefore, one should
carefully scrutinize the central assumption and its criticisms. The use of the
mathematical machinery of quantum mechanics in the proof can cause the
suspicion that `peculiar' factors that are legitimate in QM, but not justified
in a hidden variable theory, might have entered the proof. Hence, I aim to
recast Neumann's proof in a way that does not use the mathematical tools of
QM. In a way, I want to bypass the question whether Neumann was justified in
making the assumption of linear additivity of the expectation values for
hidden variable theories, by proving the \emph{unconditional validity of the
contested assumption in all physical theories}.

\section{A Straight Path to the Impossibility Proof}

Whether or not the JvNP is general, the central assumption in the proof is
valid as a fact in quantum mechanics, which is the only theory that is known
to be consistent with the empirical facts. It is also valid in classical
mechanics. How come the allegedly limited assumption made by Neumann is valid
in the only physical theories that stand today?! It might be a fact that
Neumann did not prove what he should have, but if the linear additivity is a
physical constraint to be obeyed by any physical theory, then the JvNP is
automatically valid for all theories that qualify as physical theories. I
will now show that a transparent proof of the impossibility of hidden variable
extensions of quantum mechanics can be built entirely on the necessary
fundamental physical features of microscopic mechanics, without the aid of the
mathematical features specific to quantum mechanics. There are certain
physical principles that every valid theory should obey, irrespective of the
structure and details of the theory. Among them are the fundamental
conservation laws that are related to basic space-time symmetries. However,
for microscopic physics, these can be valid only on the average and not for
individual observations in an ensemble. Yet, the requirement that the
expectation values in any ensemble obey the relations familiar in classical
mechanics is a strict physical requirement, independent of any particular
theory. Any theory where this is violated (for expectation values) does not
respect the fundamental conservation laws and space-time symmetries, and does
not qualify as a physical theory. \emph{The average value of a quantized
observable in an ensemble corresponds to a fundamental physical quantity that
is identical in meaning and scope to the same quantity in classical
mechanics}. Such quantities have to necessarily obey the physical constraints
that cannot be obeyed by single quantized values of the quantity; this is the
key point. There are other contexts where such physical constraints are
evident, like the Bohr correspondence principle and the Ehrenfest theorem.
This physical requirement is the reason for the general validity of the linear
additivity of the expectation values, irrespective of the nature of the theory
and whether the physical quantities are measurable without disturbing the
values of each other.

Consider the ensemble denoted by $\psi_{i}$, which is an eigenstate of some
observable $S$ with the eigenvalue $s_{i}$ in QM. This is `pure' and
dispersionless for $S$, with $Exp(S)=s_{i}$ but it is not dispersionless for
another observable $R$ that cannot be simultaneously measured with $S$ (I use
the term `dispersionless' for denoting a single observable that has zero
variance, and the more general term `dispersion-free' to denote ensembles with
zero variance in all observables). However, the hidden variable theories
propose that dispersion-free subensembles of the form $\left(  s_{i}%
,r_{j}\right)  $, $j=1...n$ exist. In the terminology of QM, if $\phi_{j}%
,\phi_{k}...$ are the eigenstates of another observable $R$ that is not
simultaneously measurable with $S$, the set $\left(  \psi_{i},\phi_{j}\right)
,\left(  \psi_{i},\phi_{k}\right)  ...$ defines the dispersion-free
subensembles in a hidden variable theory. $M=f(S)+g(R)$ is another valid
observable linked to $S$ and $R$ through a functional relation; for example,
$M=aR+bS$. Here, $M$ is directly measurable as a physical quantity, without
addressing the values of $R$ and $S$. Now, the crucial point is that each of
these symbols represents a physical quantity. Therefore, even though a single
value $s_{i}$ or a set of values $\left(  s_{i},r_{i},m_{i}\right)  $
notionally associated with a single copy of the physical system (particle) is
quantized, their average values in any ensemble should obey the constraints
like the conservation laws, exactly as they do in classical mechanics. Whether
or not the quantities are commuting is irrelevant for the conservation laws of
the average quantity in every individual ensemble. Therefore, the functional
relation between the quantities is maintained intact when averages over any
ensemble are taken, even though a single set of quantized values cannot obey
the relation.

This can be clarified with an example. If $R\equiv L_{x}$ and $S\equiv L_{y}$,
and $T\equiv L_{z}$, the components of the angular momentum in the directions
$x$ and $y$, and $z$, then
\begin{equation}
\vec{L}\equiv\hat{n}L=\hat{x}L_{x}+\hat{y}L_{y}+\hat{z}L_{z}=\hat{x}\sin
\theta\cos\varphi L+\hat{y}\sin\theta\sin\varphi L+\hat{z}\cos\theta L
\end{equation}
which is an exact geometrical relation that reflects the physical constraint
on the angular momentum. The operator relation in QM follows this universal
physical constraint, and it is not merely `a peculiarity of quantum
mechanics'. The measurable values of these components are quantized as $\pm1$
for each of these quantities ($L,L_{x},L_{y},L_{z}$) in a single measurement.
But, the total of these quantized values in any subensemble, like $\sum l_{i}%
$, is the total angular momentum of that ensemble in a specific direction
(figure 1). Then, it has to obey the physical law of the addition of the angular
momentum and the conservation constraint, with $\sum l_{xi}$ etc. as the
component of the angular momentum along the different directions. Thus,
\emph{every physically valid subensemble should unconditionally obey the
relation }$\sum l_{i}=a\sum l_{xi}+b\sum l_{yi}+c\sum l_{zi}$\emph{, even
though a set of single measurements obviously cannot obey }$l_{i}%
=al_{xi}+bl_{yi}+cl_{zi}$. Therefore, the relation for the linear additivity
of their expectation values $\left\langle L\right\rangle =a\left\langle
L_{x}\right\rangle +b\left\langle L_{y}\right\rangle +c\left\langle
L_{x}\right\rangle $ follows.%

\begin{figure}
	\centering
	\includegraphics[width=0.75\linewidth]{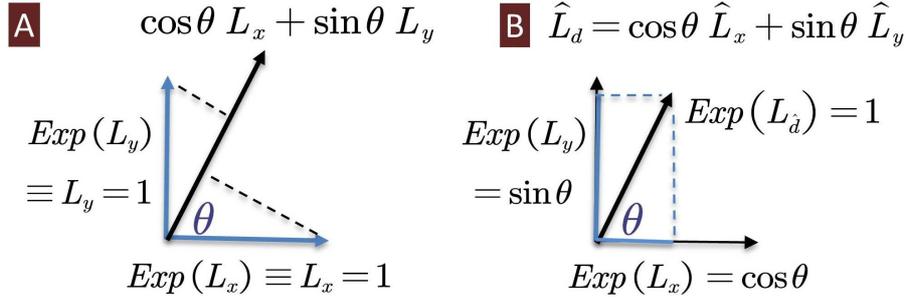}
	\caption{The linear additivity of the expectation values is an unconditional
		physical requirement on all valid theories. This example illustrates this fact
		for the components of the quantized angular momentum in two dimensions. A) The
		ensemble notionally characterized by $L_{x}=\left\langle L_{x}\right\rangle
		=+1$ and $L_{y}=\left\langle L_{y}\right\rangle =+1$. The quantity
		$L=\cos\theta L_{x}+\sin\theta L_{y}$ also has the observable values $\pm1$,
		with the maximum possible average value $+1$. But, $\left\langle
		L\right\rangle $ being an angular momentum it has to obey the linear
		additivity relation $\left\langle L\right\rangle =\cos\theta\left\langle
		L_{x}\right\rangle +\sin\theta\left\langle L_{y}\right\rangle \geq1$.
		Therefore, this is an unphysical ensemble. B) A physical ensemble necessarily
		has the right dispersion to satisfy the linear additivity relation.}
	\label{fig:linadd}
\end{figure}

In the particular example,
\begin{equation}
\left\langle L\right\rangle \equiv\sin\theta\cos\varphi\left\langle
L_{x}\right\rangle +\sin\theta\sin\varphi\left\langle L_{y}\right\rangle
+\cos\theta\left\langle L_{z}\right\rangle
\end{equation}
Since $\left\langle L_{x}\right\rangle =\sin\theta\cos\varphi\left\langle
L\right\rangle $ etc., we have the identity
\begin{equation}
\left\langle L\right\rangle =\left(  \sin^{2}\theta\cos^{2}\varphi+\sin
^{2}\theta\sin^{2}\varphi+\cos^{2}\theta\right)  \left\langle L\right\rangle
\end{equation}
which confirms the general validity of the linear additivity of the
expectation values of observables that are not simultaneously measurable. An
ensemble that violates this cannot be a physically valid ensemble because it
violates the addition and conservation of the angular momentum. We have proved
the unconditional validity of Neumann's central assumption, the linear
additivity of the expectation values,
\begin{equation}
Exp(M)=Exp(aR+bS)=aExp(R)+bExp(S) \label{linadd}%
\end{equation}
Then, it is obvious that \emph{dispersion-free ensembles are not possible}. If
an ensemble is dispersionless in the observables $R$ and $S$, it has to have a
dispersion in $M$ such that the relation \ref{linadd} is true on the average
(figure 1). Therefore, dispersion-free ensembles are unphysical and they
cannot exist in the physical world. This completes the proof of the
impossibility of hidden variable extensions of quantum mechanics.

The impossibility of dispersion-free subensembles can be demonstrated in
another way as well. We know that the subensembles can be physically separated
on the basis of the distinct values of an observable $R$ (the different $\phi
$-states), without any measurement of either $R$ or $S$. But, empirical facts
demand that if the subensembles are physically separable, then each of the
homogeneous ensembles specified by $\phi_{j},\phi_{k}...$ etc. is not
dispersion-free, which is also demanded by the relation $\phi_{i}=\sum
c_{n}\psi_{n}$. Each $\phi_{j}$-subensemble shows dispersion in the quantity
$S$, if measured. A hidden variable theory can accommodate this requirement
only by postulating a redistribution of the original $\psi_{i}$-state into
different $\psi_{n}$ states \emph{in the process of the separation} into the
pure $\phi_{j}$-subensembles. But that is not consistent with another physical
fact that no dispersion will be seen in the quantity $S$ when the different
$\phi_{j}$-subensembles are \emph{recombined before a measurement of} $S$!
Therefore, each state $\psi_{i}$ or $\phi_{i}$ represents a homogeneous
ensemble, but it is not dispersion-free. This proves that
dispersion-free ensembles do not exist.

An obvious example is when the pure state $\psi_{i}$ represents the
$x$-component of spin, $S_{x}$. An observable that is not simultaneously
measurable (noncommuting in QM) is the $z$-component of spin $S_{z}$.
Considering only these two directions, the dispersion-free ensembles in a
hidden variable theory are $\left(  +x,+z\right)  $ and $(+x,-z)$. Separating
the $+x$ ensemble into these subensembles is easy, by passing through a
$z$-directed magnetic field with a gradient. Now, there are two possibilities
before any measurement verifies the predictions: 1) the act of separation
might have perturbed the $x$-component to randomize into two ensembles $+x$
and $-x$ in each $z$-subensembles, or 2) the $x$-component is unaffected. The
second option is ruled out because it does not agree with the empirical fact
that both the $+z$ ensemble and $-z$ ensemble have the maximal dispersion in
the observable $S_{x}$. Therefore, a hidden variable theory demands the
randomization of the $x$-component in the two subensembles that are
dispersion-free in $S_{z}$. But, this also is in conflict with empirical
facts, because recombining the two $z$-subensembles before any actual
measurement of the $x$-component shows that the whole (combined) ensemble is
in the state of pure $+x$ for $S_{x}$. Therefore, the states represented by $|x+\rangle$,
$|z+\rangle$, and $|z-\rangle$ are homogeneous (pure) ensembles, but they are not
dispersion-free. The notional dispersion-free subensembles $\left(
+x,+z\right)  $ and $(+x,-z)$ cannot exist.

We can consider an example that Neumann himself mentioned, of a physical system with
its energy $E(x,p)=\left(  p^{2}/2m\right)  +V(x)$. For a harmonic
oscillator in its ground state
\begin{equation}
E_{0}=\frac{p^{2}}{2m}+\frac{m\omega^{2}x^{2}}{2}%
\end{equation}
If the hidden variable theories are possible, with dispersion-free ensembles,
$\left(  \delta(p),\delta(x)\right)  $ is one such, where the $\delta
$-functions may be taken as very narrow distributions centred on zero (nearly
dispersion-free ensemble). Though a single set of sequential measurements may
return the value $0$ for the quantities $E,p,$ and $x$, the average of $E$
over any ensemble must be the zero-point energy $\hbar\omega/2$, even without
any measurement. An ensemble in which the average zero-point energy is zero
($\ll\hbar\omega/2$) is not a physical ensemble! Hence, the subensemble
$\left(  \delta(p),\delta(x)\right)  $ is unphysical and cannot exist.

Another striking example from `new physics' is that of a stream of neutrinos,
prepared as a pure ensemble of electron type. The three initial hypothetical
dispersion-free subensembles of a hidden variable scenario are $\left(
v_{e},m_{1}\right)  ,\left(  v_{e},m_{2}\right)  $ and $\left(  v_{e}%
,m_{3}\right)  $. But, a subensemble tagged by the mass, say $m_{2}$, cannot
be uniquely $\left(  m_{2},v_{e}\right)  $! Instead, it has the dispersion
spread in the values $\left(  m_{2},v_{e}\right)  ,\left(  m_{2},v_{\mu
}\right)  $ and $\left(  m_{2},v_{\tau}\right)  $. In a way, the quantum
mechanical irreducible dispersion is displayed symbolically as $\left(
v_{e},m_{2}\right)  \neq\left(  m_{2},v_{e}\right)  $, reminding us of the
noncommuting nature of observables.

Now that we are guaranteed about the impossibility of a hidden variable
description of quantum mechanics, we can examine the influential criticisms
that kept alive the thoroughly unphysical proposal. Our verification that the
proof had complete general validity implies that the criticisms by Hermann and
Bell (and many others) were not valid. Further, it also implies that Bell's
famous counter-example must be flawed in a serious way. I will show in detail
that this is indeed the case.

\section{Grete Hermann's Criticism}

Grete Hermann's investigations in QM were motivated by the general philosophy
of causality and determinism in physics. Her criticism of Neumann's proof was actually a corollary of her
analysis of the indeterminism in QM, as represented in the Dirac formalism.
She examined the expression $Exp(O)=\langle \psi|\hat{O}|\psi
\rangle$ with $\psi=\sum c_{i}\psi_{i}$ and focussed on the
superposition $\sum c_{i}\psi_{i}$ as the source of indeterminism in the
theory. She set up the task clearly in her paper in 1935 \cite{Grete35},

\begin{quote}
Dirac's own presentation and interpretation of his formalism undeniably
includes indeterminism. Hence the question is only whether it is a necessary
component of his theory, such that in giving it up one loses essential
elements of the theory, or whether indeterminism can be detached from the
theory without thereby affecting its explanatory value.
\end{quote}

Thus the investigation was whether the indeterminism could be attributed to some elements other than the $\psi$-function. Her conclusion was that the possibility of \emph{additional hidden traits} in the
physical state, which would eliminate
indeterminism, was open and not excluded by the physical requirements of
Dirac's theory.

Hermann then proceeded to analyze Neumann's assumptions in his proof and wrote,

\begin{quote}
...the expectation value of a sum of physical quantities is equal to the sum
of the expectation values of the two quantities. Neumann's proof stands or
falls with this assumption.
\end{quote}

She noted that this is not shown as valid for the hidden variable theories.
After agreeing that this holds for the full ensemble characterized by the $\psi$-function,
Hermann poses the question whether this is true also for the subensembles of
the ensemble, which are further selected on the basis of some other
distinguishing feature that remains hidden (or not noticed) in the parent
ensemble. If the possibility of such subensembles is kept open, then one
cannot assume that the expectation value of a sum of two physical quantities
is equal to the sum of the expectation values of the two quantities. Already
in 1933 she noted \cite{Hermann-history},

\begin{quote}
...for ensembles of physical systems agreeing with one another besides in the
wave function also in terms of such a newly discovered trait, it has not been
shown that the expectation value function has the form $\langle \psi
|\hat{O}|\psi\rangle $ and is thus an $Exp(O)$.
\end{quote}

Therefore, she argued that not addressing this explicitly is equivalent to
assuming the validity of the linear additivity for the subensembles without
any justification. In other words, one is assuming that the complete
description is in the $\psi$-function and there can be no further
distinguishing features. But, the proof is about the impossibility of such
distinguishing features and therefore taking it as an assumption before the
proof makes the proof logically circular. If one assumes at the start that
there are no such features, then the deduction that such features are
impossible is a redundant fact. Hermann's conclusion was that one could hope
for a causal description of microscopic physics, since the impossibility proof
was found inadequate.

Of course, this criticism is no more valid, now that we understand that the
linear additivity is applicable to all physical ensembles. Therefore,
Neumann's assumption did not exclude any ensembles that are more general than
ensembles specified by $\psi$-functions. Neumann's assumption excluded only
unphysical ensembles, as it should.

\section{John Bell's Criticism and His Counter-Example}

John Bell's critique starts on the same point noticed by Hermann, that Neumann
did not justify his assumption of linear additivity for the hidden variable
theories when the observables were not simultaneously measurable (noncommuting
in QM). To make sure that we are discussing exactly what Bell meant, let me
quote his clear statement \cite{Bell-Fermischool},

\begin{quote}
That the statistical averages should then turn out to be additive is really a
quite remarkable feature of quantum mechanical states, which could not be
guessed \textit{a priori. }It is by no means a `law of thought' and there is
no \textit{a priori} reason to exclude the possibility of states for which it
is false. It can be objected that although the additivity of expectation values
is not a law of thought, it \emph{is} after all experimentally true. Yes, but what we
are now investigating is precisely the hypothesis that the states presented to
us by nature are in fact mixtures of component states which we cannot (for the
present) prepare individually. The component states need only have such
properties that ensembles of them have the statistical properties of observed states.
\end{quote}

Bell chose to be harsh and dismissive about Neumann's proof. He had his
reasons for this. Bell demonstrated a hidden variable model of the
measurements on a spin-1/2 particle, explicitly countering Neumann's proof
that such theories are impossible. The model had the $\psi$-function
description of the quantum state polarized along direction $\hat{a}$,
supplemented by a vector hidden variable that took random directions $\hat{r}$
in each realization of the measurement. A simple rule
\cite{Bell-RMP,Bell64,Mermin} prescribed the predictable (causal and
deterministic) discrete eigen results of $\pm1$, when measured along a
direction $\hat{n}$, given $\hat{a},\hat{n}$ and $\hat{r}$. When averaged over
the hidden variable, the correct expectation value of $\hat{a}\cdot\hat
{n}=\cos\theta$ was reproduced. This was presented as a convincing proof that
the JvNP was flawed.

Then Bell went on to examine the case of the singlet composite of two spin-1/2
particles and showed that there is a significant difference for the prediction
of spin correlations in a hidden variable theory and in QM. Again, it has been
taken for granted that the hidden variable theories discussed by Bell are
otherwise physically valid and deserved consideration as viable physical
theories. Because of this, people took up the task of testing whether the
Bell's inequalities on correlations follow the prediction of the hidden
variable theories or of QM. Finally, the results disfavoured the hidden
variable theories. The interesting point here is that if the JvNP had been convincing
as generally valid, none of those tests would have been considered worthwhile
or supported as a genuine activity in physics. However, Bell's criticism and the
successful demonstration of his counter-example eclipsed the JvNP, and created the
impression of a legitimate undecided territory that required empirical
scrutiny for the final judgement.

I have already shown that the linear additivity postulate is generally valid
in all theories that qualify as physical theories. Thus, Bell's assertion that
it ``need not be satisfied by the values of physical
quantities of hidden-variables theories, even though the grand average over
all the dispersion-free ensembles will satisfy the postulate''
is not correct. Note that nobody explored the physical reason why the linear
additivity is a ``quite remarkable feature of quantum mechanical states''.
Contrary to Bell's statements, it should have been deduced \emph{a priori}
from very general physical requirements. It is indeed a ``law of thought in
physics''. How can that be, and also consistent with the counter-example that
Bell demonstrated?! It turns out that Bell's hidden variable model is
fundamentally flawed, because it blatantly contradicts what is empirically
true! Further, we will see that the class of hidden variable theories that
Bell explored and many others tested are theories that do not respect the
fundamental conservation laws of physics.

\section{The Invalidity of Bell's Criticism of Neumann's Proof}

It is easy to show the invalidity of Bell's criticisms of the JvNP because
Bell's examples were similar to the ones that we already considered while
discussing the strengthened proof of Neumann's theorem. Bell was barking up
the wrong tree. Bell cited an example involving the measurements of the
components of a quantized spin in different directions. The expectation value
of a physical quantity is equal to the eigenvalue in a dispersion-free
ensemble. For a spin-1/2 particle, the eigenvalues of $S_{x}$ and $S_{y}$ are
$\pm1$. In quantum mechanics, these observables are noncommuting. If we
consider the bisecting direction, the quantum mechanical operator $S_{bs}$
involves the combination $\left(  S_{x}+S_{y}\right)  /\sqrt{2}$. Again the
eigenvalues are $\pm1$. However, for a dispersion-free ensemble, it cannot be
true that $Exp\left(  S_{x}+S_{y}\right)  =\left(  Exp\left(  S_{x}\right)
+Exp\left(  S_{y}\right)  \right)  $ because $Exp\left(  S_{x}+S_{y}\right)
=\pm\sqrt{2}$, and $Exp\left(  S_{x}\right)  $ and $Exp\left(  S_{y}\right)  $
are $\pm1$.

Bell's scathing criticism was that Neumann made the
``silly'' assumption that the linear additivity of the
expectation values should be obeyed by the hidden variable (dispersion-free)
ensembles as well, just like the quantum mechanical ensemble. However, here
Bell made a fundamental mistake. I repeat the proof discussed earlier,
transcribed to Bell's example.

The quantity $Exp\left(  S_{x}+S_{y}\right)  /\sqrt{2}$ is the \emph{average}
spin angular momentum in the direction that bisects $\hat{x}$ and $\hat{y}$.
It is a physical quantity, averaged over the ensemble. \emph{In any other
direction, the components of the angular momentum should obey the law of vector projection, expected in any physical theory}, without regard to the commutation
rules, discreteness of spin projections etc. That is, though the spin
projection for a single particle is $\pm1$ in any direction, the average over
any statistical ensemble is the same as the average angular momentum, which
can take any value between $-1$ and $+1$. Then, necessarily, it is the
vector sum of the average angular momentum in the directions $\hat{x}$ and
$\hat{y}$, $Exp\left(  S_{x}\right)  +Exp\left(  S_{y}\right)  $. Thus,
\begin{equation}
Exp\left(  aS_{x}+bS_{y}\right)  =aExp\left(  S_{x}\right)  +bExp\left(
S_{y}\right)
\end{equation}
This is a physical truth, to be obeyed in any valid theory. Ensembles that do
not obey this cannot exist in the physical world! Then, Neumann's proof is
resurrected without a blemish: dispersion-free ensembles are impossible
because all physical ensembles obey the linear additivity of the expectation
values, as Neumann assumed.

We can go through the exercise of dividing the ensemble corresponding to the
state into dispersion-free ensembles. Let the state be $\left\vert
n\right\rangle $, polarized along $\vec{n}=\hat{x}+\hat{y}$. Then the values
along $\hat{n}$ are all $+1$. The average spin angular momentum of the state
is $+1(\hbar/2)$ along $\hat{n}$. The dispersion-free ensembles of possible
values for $S_{x}$ and $S_{y}$ are $\left(  +1,+1\right)  ,\left(
+1,-1\right)  ,\left(  -1,+1\right)  ,\left(  -1,-1\right)  $. The average
angular momentum of these subensembles are $\sqrt{2}\hat{n},\sqrt{2}\hat
{n}^{\prime},-\sqrt{2}\hat{n}^{\prime}$ and $-\sqrt{2}\hat{n}$, where $\hat
{n}^{\prime}$ is the unit vector perpendicular to $\hat{n}$. None of these is
compatible with the conservation of the angular momentum and the value of the spin
angular momentum of the parent ensemble, which is $+1\hat{n}$. Therefore,
such ensembles are unphysical and cannot exist, exactly as Neumann asserted.

I now proceed to show that Bell's famous counter-example of a hidden variable
model of spin measurements fails in a way that is quite independent of whether
or not it obeys the assumption in Neumann's Proof.

\section{The Inconsistency of Bell's Hidden Variable Model}

As already mentioned, Bell's hidden variable counter-example, meant to
disprove Neumann's assertion, was for the measurements of the component of a
spin polarized along a direction $\hat{a}$, measured along the direction
$\hat{n}$. In QM, this is the state $\left\vert a+\right\rangle $ and the
results of the measurements are $S_{n}=\pm1$ in any direction $\hat{n}$, with
the expectation value $\left\langle a+|\hat{\sigma}_{n}|a+\right\rangle
=\hat{a}\cdot\hat{n}=\cos\theta$. Bell constructed the model by supplementing
the state polarized along $\hat{a}$ with a vector hidden variable $\hat{r}$
that took random directions in each realization of the measurement. A simple
nonlinear rule prescribed the predictable eigen results of $\pm1$, when
measured along $\hat{n}$ \cite{Bell64,Bell-RMP,Mermin}. The randomness in the
observable values arose from the randomness in the hidden variable. The
average over the directions of the hidden variable reproduced the correct
expectation value $\hat{a}\cdot\hat{n}=\cos\theta$. We take this apparently
successful model and analyze the difference between QM and the hidden variable
theories. Consider the simple case when $\hat{a}=\hat{n}=\hat{x}$. Then the
result for each particle is $S_{x}=+1$, and the expectation value is
$Exp(S_{x})=+1$. In the hidden variable theory, each particle in the total
ensemble has $S_{x}=+1$, with $S_{y}$ and $S_{z}$ distributed as $\pm1$
equally. In contrast, in QM, each particle is in the state $\left\vert
x+\right\rangle $ with $S_{x}=+1$, which is also an equal superposition of the
states $\left\vert y+\right\rangle $ and $\left\vert y-\right\rangle $, and
also of $\left\vert z+\right\rangle $ and $\left\vert z-\right\rangle $.
\emph{It is on this crucial difference that Bell's famous model collapses}.
This can be demonstrated in many ways.%

\begin{figure}
	\centering
	\includegraphics[width=0.95\linewidth]{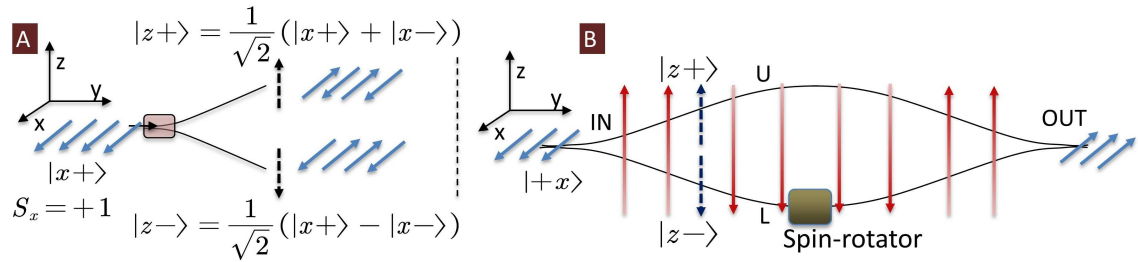}
	\caption{The failure of Bell's hidden variable counter-example to Neumann's
		impossibility assertion. A) The initial polarization of the spin-1/2 particles is
		in the $+x$ direction. A small magnetic field with a constant gradient acts as
		an ensemble separator for spin projections $+1$ and $-1$ in the $z$-direction.
		QM predicts the detection of equal proportions of $+x$ and $-x$ polarization
		in each subensemble! B) A symmetric Stern-Gerlach interferometer between the
		state preparation and detection. The spin rotator (SR) in only the lower path
		rotates the spin by $2\pi$ about the $x$-axis. With the SR absent, there is no
		change in the spin state. With the SR, all the particles are detected with the
		spin state reversed, in the $-x$-direction. A hidden variable theory does not
		reproduce this result.}
	\label{fig:figure-sg}
\end{figure}

As indicated in the figure 2(A), we introduce an ensemble separator that
divides the particles into a subensemble that would give $S_{z}=+1$, if
measured, and another with $S_{z}=-1$. A spatially limited magnetic field $B$
in the $z$-direction, with a constant gradient $B^{\prime}$ will do this job.
In the hidden variable model, the $S_{x}=+1$ parent ensemble is divided
further into two subensembles of $S_{z}=+1$ and $S_{z}=-1$, in equal
proportions. The magnetic field would have perturbed the $x$-component to
precess by some amount, which depends on the exact value of the field and the
duration for which the particle is in the field. But in QM, each particle in
the $\left\vert z+\right\rangle $ state (or the $\left\vert z-\right\rangle $
state) is an \emph{equal superposition} of $\left\vert x+\right\rangle $ and
$\left\vert x-\right\rangle $, irrespective of the field used in the ensemble
separator. Clearly, Bell's hidden variable model fails completely now to
reproduce the results of the measurements in the subensembles, in the $x$-direction.

Another dramatic demonstration of the inadequacy of Bell's model is when an
active version of the Stern-Gerlach (S-G) closed interferometer is inserted
between the state preparation and the measurement apparatus (figure 2(B).
There is a spin-rotator in one of the arms (say, lower arm), \emph{which
rotates the spin by }$2\pi$\emph{ about the }$x$\emph{-axis}. Hence, \emph{the
}$x$\emph{-polarization is not affected by the spin rotator}. Both the $y$ and
$z$ components rotate through full $360^{o}$. When the spin rotator is
not operative, the particles polarized along $\hat{x}$ enter the symmetric
device and come out without any change in the polarization state. This is
certainly so in QM, and let us assume that similar results are obtained in
Bell's model as well (though it is not the case in a detailed scrutiny). When
the spin rotator is operative, nothing should change in the hidden variable
model: the $x$-polarization is not touched and the $z$ and $y$ polarizations
are restored back after the full $2\pi$ rotation. But, the QM result is very
different. The state that enters the S-G device is
\begin{equation}
\left\vert x+\right\rangle =\frac{1}{\sqrt{2}}\left(  \left\vert
z+\right\rangle +\left\vert z-\right\rangle \right)
\end{equation}
The state that exits after the spin rotator acts on the $\left\vert
z-\right\rangle $ state is
\begin{equation}
\left\vert s\right\rangle =\frac{1}{\sqrt{2}}\left(  \left\vert
z+\right\rangle +e^{-2\pi/2}\left\vert z-\right\rangle \right)  =\frac
{1}{\sqrt{2}}\left(  \left\vert z+\right\rangle -\left\vert z-\right\rangle
\right)  =\left\vert x-\right\rangle
\end{equation}
Then \emph{all the particles should be measured as }$S_{x}=-1$, with the
expectation value $Exp(S_{x})=-1$, though all the particles entered with
$S_{x}=+1$! The failure of the hidden variable counter-example is total. So,
Bell's hidden variable scheme for single spin fails, contrary to the widely
held general impression. The quantized spin component is only one of the
features of a `state' in quantum mechanics; the evolution of the relative
phase of the states, which remains alien to such hidden variable theories,
determines the measured values and their distribution.

All these results are explicit demonstrations of the fact that a hidden
variable description of even simplest quantum mechanical situations is
inconsistent. The dispersion of quantum mechanics is due to some form  of
wave-particle duality that cannot be reduced to the trivial statistical
dispersion of deterministic hidden variables. Having proved the unphysical
nature of Bell's hidden variable model for single spin, I now proceed to show
the drastic unphysical nature of the local hidden variable theories (LHVT) of
two-particle states, for which Bell derived the Bell's inequality and prompted
experimental tests \cite{Unni-EPL,Unni-Pramana}. In fact, those hidden
variable theories had no physical sanity to start with and it was easy to rule
them out as unworthy of any experimental test. Surely, people who spent
enormous efforts in testing the LHVT would not have done so if they had
realized that they were testing thoroughly unphysical theories that grossly
violate the conservation of the angular momentum. In that sense, this result is
gloomy, but its presentation here is essential.

Before we discuss the proof of the unphysical nature of the LHVT, I note that
Bell's exercise was motivated by the desire to study whether it was possible to cure the 
`incompleteness' of QM, pointed out in the
Einstein-Podolsky-Rosen (EPR) paper in 1935 \cite{EPR}, using hidden variables. Bell had obviously thought
that the `incompleteness' mentioned in the EPR paper was the same as the `incompleteness' discussed by Neumann, the lack of deterministic
causality, for which the prescription was a local hidden variable theory. He
stated as much in the opening paragraph of his 1964 paper. However, the
`incompleteness' in the EPR paper is very
different from the `causal incompleteness'
addressed by Neumann, Hermann, and Bell! Since this is a crucial point for
physics, I prove this point before showing the inherently unphysical nature of
the LHVT.

\section{The Confusion of Two Kinds of `Incompleteness'}

We expect that physicists are careful in distinguishing two entirely different
conceptual points for which the same term has been used, because physics has
the advantage of the availability of precise mathematical representations for
the statement of each important concept. However, in the case of quantum
mechanics an unfortunate `uncertainty' has crept in. I am referring to the
indiscriminate use of the term 
`incompleteness' to designate two different notions in the
discussion of QM! It should be obvious to most that Neumann's use of that term
in 1932, in the context of causality, couldn't be in the same sense as
Einstein's use in 1935 in the context of physical reality; otherwise Einstein
would not have bothered to define explicitly that term in the EPR paper and
beyond. To be (linguistically) precise, at least at this late stage in the
situation, let me quote Neumann in detail, in his first use in the context of
indeterminism in QM:

\begin{quotation}
If we want to explain the non-causal character of the connection between
$\psi$ and the values of physical quantities following the pattern of
classical mechanics, then this interpretation is clearly the proper one: In
reality, $\psi$ does not determine the state exactly. In order to know this
state absolutely, additional numerical data are necessary. That is, the system
has other characteristics or coordinates in addition to $\psi$. If we were to
know all of these we could then give the values of all physical quantities
exactly and with certainty... It is customary to call these hypothetical
additional coordinates ``hidden parameters''
or ``hidden coordinates'', since they must
play a hidden role, in addition to the $\psi$ which alone have been uncovered
by investigation thus far. Explanations by means of hidden parameters have (in
classical mechanics) reduced many apparently statistical relations to the
causal foundations of mechanics. An example of this is the kinetic theory of gases.

Whether or not an explanation of this type, by means of hidden
parameters, is possible for quantum mechanics is a much discussed question.
The view that it will sometime be answered in the affirmative has at present
some prominent representatives. If it were correct, it would brand the present
form of the theory provisional, since then the description of states would be
essentially incomplete.
\end{quotation}

What I just quoted from Neumann's treatise is the `determinism
incompleteness' or `causality incompleteness' of QM, which was very much the
part of debates right from Heisenberg's discussion of the uncertainty
relations in 1927. In fact, Hermann's criticism of Neumann's proof was in the
context of her investigations of the question of causality.

Now, I quote from the EPR paper \cite{EPR}, their definition of ``completeness'' of a theory:

\begin{quote}
...every element of the physical reality must have a counterpart in the
physical theory We shall call this the condition of completeness. 
\end{quote}

Thus, the term ``completeness'' in the sense of EPR demands that there is a
one-to-one correspondence between the physical states and the QM states
represented by the $\psi$-functions. \emph{The EPR paper does not mention the
terms `indeterminism' or `uncertainty', or even `causality'}, anywhere. Nor
does it mention hidden variables and such. In fact, the EPR claim, quoting
from the paper is

\begin{quote}
Previously we proved that either (1) the quantum-mechanical description of
reality given by the wave function is not complete or (2) when the operators
corresponding to two physical quantities do not commute the two quantities
cannot have simultaneous reality.
\end{quote}

Note carefully that the second option is the denial of hidden variable
theories, which is the one EPR opted for! Continuing with their proof (italics
are mine),

\begin{quote}
\emph{Starting then with the assumption that the wave function does give a
complete description of the physical reality, we arrived at the conclusion
that two physical quantities, with noncommuting operators, can have
simultaneous reality}. Thus the negation of (1) leads to the negation of the
only other alternative (2). \emph{We are thus forced to conclude} that the
quantum-mechanical description of physical reality given by wave functions is
not complete.
\end{quote}

It is stated clearly that the authors (EPR) do not accept the negation of
the option (2); they were `forced' to accept the option (1) because of their
conviction in the impossibility of simultaneous values for noncommuting
observables. Thus, there is not even the slightest ambiguity that the EPR
assertion of `incompleteness' followed from denying dispersion-free ensembles
and simultaneous values of noncommuting observables -- a situation that
accepts the acausal nature of QM. The `EPR-incompleteness' means that there is
no one-to-one correspondence between the physical states and their
representation by the $\psi$-functions of QM. Therefore, it is amply clear
that the `representation incompleteness' in the EPR paper, which demanded an
entirely different theory, is not at all the same as Neumann's `determinism
incompleteness', proposed to be cured with a hidden variable theory.

Yet, it is imperative that I explicitly prove that the `EPR-incompleteness' is
different from the `determinism incompleteness' that motivated the hidden
variable program. There are many reasons for this. One sufficient reason is
the stark contrast and conflict between the statements quoted from the EPR
paper and the opening sentence of Bell's 1964 paper titled ``On the
Einstein-Podolsky-Rosen paradox'':

\begin{quote}
The paradox of Einstein, Podolsky and Rosen was advanced as an argument that
quantum mechanics could not be a complete theory but should be supplemented by
additional variables.
\end{quote}

Try as one might, no mention of what Bell says can be found in the EPR paper,
that ``QM should be supplemented by additional variables'', even as an
implication! This point has been mildly raised by A. Fine earlier
\cite{Fine}. It is very plausible that Bell really did not consult the original EPR paper, but relied on a paper by D. Bohm and Y. Aharonov in 1957 that discussed the EPR situation in terms of the spin variables and nonlocal (superluminal) physical signals \cite{Bohm-EPR}. This is suggested by certain facts. Bohm and Aharonov made an error in the order of the authors in their citation to the original article, as Einstein-Rosen-Podolsky (ERP) instead of Einstein-Podolsky-Rosen (EPR).\footnote{In fact, the citation is unusually casual, without mentioning the initials of the three authors. The scrambled order can be traced back to Bohm's textbook, ``Quantum Theory'', published in 1951 \cite{Bohm-QM}.} Then they called the EPR argument as the ``ERP paradox''. Bell repeated this error of bibliographic citation and also the description of the EPR argument as a `paradox', in his 1964 paper, which is a curious correlation. In any case, the proof that the `EPR-incompleteness' is very
different from the `causal incompleteness' that Bell refers to is easy.
Considering the importance of such a proof, let us go through the argument in
the EPR scenario of two particles that had a prior interaction and then
separated into two spatially distinct regions A and B.

Assume that the randomness in the observed values is only apparent and a more
complete hidden variable description exists. As EPR and Bell did, we assume
strict Einstein locality. Now consider the joint $\psi$-function for the two
particles, $\Psi_{12}$.  The hidden variable description consists of this
function and a set of hidden variable realizations at the locations A and B.
There is no causal indeterminism left in this hypothetical scenario because we
have already admitted the hidden variables. Now I show that the `EPR-incompleteness' is still present! A measurement at the location A results in a
new localized $\psi$-function $\psi_{1}$ at A and another correlated function
$\psi_{2}$ at the location B. \emph{Therefore, a measurement at A did change
the }$\psi$\emph{-function representation of the physical state at B, from
}$\Psi_{12}$\emph{ to }$\psi_{2}$, whether or not the hidden variables are
operative. However, we have assumed strict locality, that the physically
factual state at B cannot be changed by a measurement at A. Since $\Psi_{12}$
and $\psi_{2}$ represent two different physical states, \emph{it is proved
that there is no one-to-one correspondence between the }$\psi$\emph{-function
representations of the theory (supplemented by the hidden variables) and the
factual physical states}. We have proved the `EPR-incompleteness' even for a
hypothetical hidden variable deterministic extension of QM! This is a
remarkable result that shows how far unrelated is the EPR argument from the
hidden variable issue.

We are now in a quandary, because Bell and numerous others mixed up and
equated the two unrelated notions of incompleteness, and proceeded with their
assertions. Einstein warned about such a misunderstanding when he commented on
M. Born's (mis)understanding of the EPR argument \cite{Born}:

\begin{quote}
The whole thing is rather sloppily thought out, and for this I must
respectfully clip your ear. I just want to explain what I mean when I say that
we should try to hold on to physical reality... But whatever we regard as
existing (real) should somehow be localised in time and space. That is, the
real in part of space A should (in theory) somehow `exist' independently of
what is thought of as real in space B. When a system in physics extends over
the parts of space A and B, then that which exists in B should somehow exist
independently of that which exists in A. That which really exists in B should
therefore not depend on what kind of measurement is carried out in part of
space A; it should also be independent of whether or not any measurement at
all is carried out in space A. If one adheres to this programme, one can
hardly consider the quantum-theoretical description as a complete
representation of the physically real.
\end{quote}

Einstein was always referring to `reality' in the sense of `objective
physical existence' and not in the sense of deterministic values for all
physical observables. The notion of `local realism' implied in the EPR paper
is very different from the goal of `local determinism' in a hidden variable
theory. This has been confirmed by the ever-careful W. Pauli, in his revealing letters to M. Born in 1954 \cite{Born}, when he assumed the role of an adjudicator in the friendly debate between Born and Einstein. Pauli wrote, 
\begin{quote}
	It seemed to me as if you
	had erected some dummy Einstein for yourself, which you then
	knocked down with great pomp. In particular, Einstein does
	not consider the concept of `determinism' to be as fundamental
	as it is frequently held to be (as he told me emphatically many
	times)... In the same way, he disputes that
	he uses as criterion for the admissibility of a theory the question:
	`Is it rigorously deterministic?'
	Einstein's point of departure is `realistic' rather than `deterministic',
	which means that his philosophical prejudice is a
	different one... it
	seems to me misleading to bring the concept of determinism
	into the dispute with Einstein.
\end{quote}
In a follow up letter he wrote to Born, ``I have already tried in my last letter to explain Einstein's
point of view to you. It is exactly the same in Einstein's
printed work and in what he said to me.'' This shows that Pauli's sharp and straight criticism applies  to Bell and many others, who misread between the lines in `the printed work', and failed to notice that Einstein's point of departure was `realistic' rather than `deterministic'.

Stated in the context of the often quoted Einstein-anecdote of the
``reality of the moon when one is not looking'', the `EPR-completeness' insists on
the objective physical existence of the matter-moon, but does not demand
anything about its trajectory that is more deterministic than what is
specified with an uncertainty $\hbar$ of action. It is really an unfortunate
turn of the course of physics that this kind of `EPR-incompleteness' was widely
confused as the `determinism incompleteness' requiring hidden variables. The
transference of a linguistic confusion to a lasting physical delusion with
myriad irrational features like telepathic nonlocality is a serious issue.
Einstein was vocal and explicit in his conviction that the $\psi$-function is
already the representation of the statistical ensemble, and not of the single
system. Then, it is obvious that supplementing $\psi$ with any number of
hidden variables cannot restore determinism and causality. Einstein would have
clipped our ears, and not so respectfully, for confusing his proof of
`representation incompleteness' as a demand for a quantum theory supplemented
with the hidden variable dressing.

\section{The Unphysical Nature of the LHVT}

The spin correlation of a spin-singlet concerns the measurements on two
(spin-$1/2$) particles with the total spin zero, and the possible results at
locations A and B are $A_{i}=\pm1$ and $B_{i}=\pm1$. When both the
measurements are in the same direction, the angle between the apparatus
directions is zero, $\theta=0$. Then, if $A_{i}=+1$, then $B_{i}=-1$ and vice
versa; only then the total spin is zero, as demanded by the conservation of
the angular momentum. To see what \emph{the expected correlation is for a
general angle, dictated by just the conservation of the angular momentum},
consider the measurements at A and B with the difference $\theta\neq0$ in the
angular settings. The correlation is the average of a large number of products
$A_{i}B_{i}$.

Since the average of a sum of quantities is the same as the sum of the
averages of two subsets that are half the size, we can make the subsets
$A_{i}(+1)B_{i}$ and $A_{i}(-1)B_{i}$, where the first set has all $A_{i}=+1$
and the second has all $A_{i}=-1$. Then the correlation function is
\[
C_{e}(\theta)=\frac{1}{N}\sum A_{i}B_{i}=\frac{1}{N/2}\sum A_{i}%
(+1)B_{i}+\frac{1}{N/2}\sum A_{i}(-1)B_{i}%
\]

This involves only a simple reordering of the pairs of data, \emph{keeping the
pairwise data intact}, which does not affect any average (we can do the same
exercise with the B values as the anchor -- the situation is symmetric). Now
note that the first set has the average angular momentum at A as $+1$ and the
second set as $-1$ (in units of $\hbar/2$). Then, the conservation of the
angular momentum on the average dictates that the corresponding average value
at B, at a relative angle $\theta$, should be the projection of the opposite
angular momentum along the direction of B; in the first set it should be
$-\cos\theta$ and in the second, $+\cos\theta$. Then the average for the whole
set is just
\begin{equation}
C_{e}(\theta)=\frac{1}{2}\left[\left(  +1\times-\cos\theta\right)  +\left(  -1\times+\cos
\theta\right)\right]  =-\cos\theta
\end{equation}
\emph{This is predicted to be the experimentally observable singlet
correlation, purely from the conservation of the angular momentum, independent
of any theory}. I have not even mentioned QM in this derivation. Sure, the
prediction of QM agrees with this because it is a theory that respects the
conservation laws for the average over the ensembles. A theory that has a
different prediction is then definitely not compatible with the conservation
of the average angular momentum. Obviously, such theories are unphysical.
\emph{Since the LHVT predict a very different correlation function that has a
linear dependence on }$\theta$\emph{ for small angles, the LHVT are in that
unphysical class}. They grossly violate the fundamental conservation laws. It
is unlikely that anyone would have tested them, or even discussed them, if
this result had been known. But this result can be seen as an extension of the
physical factors that resurrect Neumann's proof, which was always known.

\section{Bohm's ``Hidden Variable'' Quantum Mechanics}

When D. Bohm published his reinterpretation of the Schr\"{o}dinger quantum
mechanics \cite{Bohm-PR}, he preferred to call it a ``hidden
variable'' construction. This was because the position
variable in the $\psi$-function was treated as a hidden variable. However, it
is very important to emphasize that \emph{the theory was based on just a
reinterpretation of the Schr\"{o}dinger equation of standard quantum
mechanics, without any modification}. All the peculiarities related to the
$\psi$-function ($\psi(x,t)=A\exp(iS/\hbar)$) were transferred to the nonlocal
quantum potential, $\left(  \nabla^{2}A\right)  /A$. Therefore, it was
guaranteed to give the same statistical results as the Schr\"{o}dinger quantum
mechanics, while allowing notional trajectories, albeit in the simplest cases
involving single particle space-time quantum mechanics. But, the theory has
many unphysical features and inadequacies. As it is well known, the special
status given to the position variable has not allowed any viable treatment of
most of the standard quantum mechanical problems involving spin, atomic
excitation and radiation, unstable particles, two-particle correlations etc.
If a quantum theory cannot deal with atomic spectra and two-particle
correlations, what relevance can it possibly have, except as a distracting
example in a discussion of myriad features of the Schr\"{o}dinger mechanics? In
any case, classifying it as a hidden variable theory of the kind Neumann
discarded with his proof is incorrect, because \emph{Neumann's proof was for
the impossibility of hidden variable theories that allowed dispersion-free
ensembles}. In Bohm's theory, there are no physical ensembles that are
dispersion-free in all possible observables, even with the `quantum potential'.

As a remark aside, the denial of clearance for Bohm to work in the Manhattan
project and his eventual exile from the USA to Brazil in 1951 prevented any
interaction between Neumann and Bohm. As for Neumann's attitude towards Bohm's
claim of a hidden variable reinterpretation of QM, a letter from Bohm to Pauli
in 1951 \cite{Bohm-Pauli} has a relevant mention: \textquotedblleft It appears
that von Neumann has agreed that my interpretation is logically consistent and
leads to all results of the usual interpretation. (This I am told by some
people.) Also, he came to a talk of mine and did not raise any
objections\textquotedblright. However, this much is clearly not an approval
for a hidden variable theory of the kind Neumann analyzed, in which \emph{the
characteristic feature is the existence of dispersion-free ensembles}. It would
have been ironical that the major work of a physicist who was denied access to
the nuclear research in the USA apparently trashed the famous proof of the
illustrious commissioner of the US Atomic Energy Commission! Though we do not
have any record of what Neumann thought about Bohm's hidden variable claim, we
have Bohm's view of Neumann's proof, for example, in the same letter:

\begin{quote}
(Neumann's) proof involves the demonstration that no 
``dispersionless'' states can exist in the quantum theory, so
that no \textit{single} distribution of hidden parameters could possible determine the
results of \textit{all} experiments (including for example, the measurements of
momentum and position). However, von Neumann implicitly assumes that the
hidden variables are only in the observed system and not in the measuring
apparatus. On the other hand, in my interpretation, the hidden variables are
in \textit{both} the measuring apparatus and the observed system. Moreover, since
different apparatus is needed to measure momentum and position, the actual
results in each respective type of measurement are determined by \textit{different}
distributions of hidden parameters. Thus, von Neumann's proof is irrelevant to
my interpretation.
\end{quote}

However, in his book ``The Undivided Universe'' \cite{Bohm-universe}, Bohm
directly relies on Bell's argument and treads tentatively in his
interpretation of Neumann's proof, almost cautioning us to the fact that the
linear additivity is nevertheless true in quantum mechanics, and in the
empirical data of measurements. Also, Bohm's well-known text book on quantum
mechanics \cite{Bohm-QM}, published first in 1951, has a section titled,
``Proof that quantum theory is inconsistent with hidden
variables''! The republished editions retain the section.
There, mentioning the necessity of wave-particle duality as well as the
analysis of the EPR argument, Bohm writes,

\begin{quote}
We conclude then that no theory of mechanically determined hidden variables
can lead to all of the results of the quantum theory.
\end{quote}
But, there is no mention of this in the sections discussing the proofs of the
impossibility of hidden variables, in the later book, ``The Undivided
Universe''. One gets the impression that Bohm considered his theory as a
`different kind of hidden variable theory', while agreeing with the general
view that \emph{the hidden variable theories cannot reproduce all the results
of the quantum theory}.

It is very easy to prove that Bohm's theory with the spatial position as a
hidden variable is no replacement for quantum mechanics. Consider a source of
electron neutrinos, like the decay of neutrons. Every neutrino in this example
is of the electron flavour, when it is emitted. We know that there is a finite
probability for the detection of the neutrino as muon type.  But, the initial
position, which is Bohm's hidden variable, or the guiding equation $v=\nabla
S/m$, is not relevant for the random detection of neutrinos in separate
flavours. In other words, the much touted determinism in the theory is an
illusion. So, Bohm's theory is just a toy model still, and not useful as a
general theory of physical phenomena. Those who are familiar with the damning
criticisms with explicit examples, like that of Einstein's in 1953
\cite{Einstein-Bohm}, would know how unphysical the theory can be. In
Einstein's example, a macroscopic particle in a box has the wavefunction
$\psi(x)=A\exp(iS/\hbar)\sim\exp\left(  -ipx\right)  +\exp\left(  ipx\right)
$ and velocity $v=\nabla S/m=0$. So, the particle remains at rest, until
observed, when it acquires the velocity $\pm p/m$ within the duration of the
observation, \emph{implying an arbitrary hidden acceleration}. This example is
seriously damaging for Bohm's theory when reconsidered with photons. This kind
of unavoidable physical inconsistency of Bohm's theory is related to its
nonlocal quantum potential. Einstein had perhaps thought through several
aspects of such a theory because his own discarded attempt in 1927, at a
causal interpretation of the Schr\"{o}dinger equation, had many similarities
to Bohm's later attempt \cite{Einstein-1927}. Further, as Bohm himself
emphasized, the poly-dimensional spatial nonlocality, inherent in the
$3n$-dimensional $n$-particle $\psi$-function, is the hallmark of the theory.
But, this unphysical aspect goes against the core tenet of relativity, as Bohm admitted in his paper on the `ERP-paradox', ``It must be admitted, however, that this quantum
potential seems rather artificial in form, besides being
subject to the criticism... that it implies instantaneous interactions between distant particles, so that it is not consistent with the theory of relativity.''

\section{What Prohibits Dispersion-Free Ensembles?}

Since it has been shown in multiple ways, and by analyzing explicit examples,
that hidden variable extensions to QM and dispersion-free ensembles cannot
exist, I will now discuss the core physical reason for the general result that
the dispersion in mutually incompatible observables is irreducible. Let an
ensemble be defined through  a $\psi$-function. All $\psi$-functions are of
the form $\psi(x,t)=A\exp(iS/\hbar)$, where the function $S$ is the action.
Usually, the QM dispersion is presented through the uncertainty principle
involving two conjugate observables, which in turn finds a justification in
the wave-particle duality. However, the real origin of the quantum uncertainty
is the intrinsic uncertainty in the action, characterized by its scale $\hbar
$; therefore the irreducible uncertainty should be correctly expressed as
$\Delta S\geq\hbar$ \cite{Unni-RQM}. Then, without any forced interpretation
of ascribing an unrealistic wave nature to the matter particles themselves, we
see that there would be the irreducible uncertainty relation between variables
that define the action, which are by nature the conjugate variables of
dynamics. The action is the product of a dynamical quantity (like the momentum
$p$) and the corresponding coordinate (displacement $x$). \emph{Since the
dispersion in the }$\psi$\emph{-ensemble is in the action, }$\Delta S\geq
\hbar$\emph{, it cannot be split into definite values of both a dynamical
variable and a conjugate coordinate}. That is, a dynamical ensemble in
mechanics is characterized by the action and its irreducible variance.
Therefore, dispersion-free ensembles do not exist. In fact, this statement is
more general, applicable to all mechanics, but I do not discuss the details
here. A full description may be found in the reference \cite{Unni-RQM}.
Briefly stated, all the foundational problems of QM can be traced to
misinterpreting the Schr\"{o}dinger equation and its $\psi$-function as
pertaining to the evolution of a single dynamical history, or a single
particle, whereas they factually describe ensemble dynamics. In fact, the
Schr\"{o}dinger equation is the same as the evolution equation of the
probability density of the ensemble, with the constraint that the ensemble
obeys Hamilton's action dynamics. This is proved in the reference
\cite{Unni-RQM}, but it is easy to briefly state what is involved.

The continuity equation for the classical probability density $\rho(x,t)$ of a
statistical ensemble concerning the dynamics of a particle is
\begin{equation}
\frac{\partial\rho}{\partial t}=-\nabla\cdot j
\end{equation}
The probability current  $j=\rho v=\rho\nabla S/m$ where $S(x,p,t)$ is the
action and $m$ is the mass of the particle. Defining the positive quantity
$\rho=\psi\psi^{\ast}$, where $\psi=Ae^{i\phi}$, the continuity equation is
\begin{equation}
\frac{\partial\left(  \psi\psi^{\ast}\right)  }{\partial t}=\psi\frac
{\partial\psi^{\ast}}{\partial t}+\psi^{\ast}\frac{\partial\psi}{\partial
t}=-\nabla\cdot j=-\frac{1}{m}\nabla\cdot\left(  \psi\psi^{\ast}v\right)
\end{equation}
If the dummy angle $\phi$ is defined such that $\nabla\phi=v$, the equation
can be written entirely in terms of $\psi,\psi^{\ast}$ and their spatial
derivatives.  We have $\nabla\psi=i\psi\nabla\phi+e^{i\phi}\nabla A$.
Multiplying by $-i\psi^{\ast}$ gives the desired current term $\rho v$ and
an additional term. Then%
\begin{equation}
-i\psi^{\ast}\nabla\psi=\rho v-i\psi^{\ast}e^{i\phi}\nabla A=\rho v-iA\nabla A
\end{equation}
Adding the complex conjugate eliminates the second term,%
\begin{equation}
i\left(  \psi\nabla\psi^{\ast}-\psi^{\ast}\nabla\psi\right)  =2\rho v=2j
\end{equation}
Therefore, the continuity equation takes the form%
\begin{equation}
\psi\frac{\partial\psi^{\ast}}{\partial t}+\psi^{\ast}\frac{\partial\psi
}{\partial t}=\frac{i}{2}\nabla\cdot\left(  \psi^{\ast}\nabla\psi-\psi
\nabla\psi^{\ast}\right)  =\frac{i}{2}\left(  \psi^{\ast}\nabla^{2}\psi
-\psi\nabla^{2}\psi^{\ast}\right)
\end{equation}
But, $\partial\psi/\partial t=\left(  i/2\right)  \nabla^{2}\psi$ is the free
particle Schr\"{o}dinger equation! With the identification of $\phi$ with the
scaled action of dynamics, $S/\hbar$, the correspondence is complete. What is
crucially important is that \emph{the }$\psi$\emph{-function describes the
statistical ensemble, and does not represent a matter-wave or the single
particle dynamics}, exactly as Einstein speculated. 

\emph{The single particle dynamics is governed by another new dynamical
equation for the }$\zeta$\emph{-function of the action }\cite{Unni-RQM},
defined as $\zeta(x,t)=\exp(iS/\hbar)$,%
\begin{equation}
\frac{\partial\zeta(S)}{\partial t}=-\frac{i}{\hbar}H\zeta
\end{equation}
This equation,\emph{ universally applicable to the dynamics at all scales of
masses and velocities}, contains the standard Hamilton-Jacobi equation and a
small additional term that reflects the intrinsic uncertainty of the action.
Thus, the correct equation for the nonrelativistic dynamics of a single
particle is a modified Hamilton's equation \cite{Unni-RQM},
\begin{equation}
\frac{\partial S}{\partial t}=-H+\frac{i\varepsilon}{2m}\nabla^{2}S
\end{equation}
In the factual quantum dynamics, there are neither matter-waves, nor pilot waves with a quantum
potential. The particle is separate from the `action-wave' $\zeta
(x,t)=\exp(iS/\hbar)$, and all the results of the single particle interference
are reproduced without the collapse of the state. There is no measurement
problem either, and full ontological consistency and locality are restored. The
two-particle correlations follow from the local interference of the
action-waves. The fundamental uncertainty $\Delta S\geq\hbar$ persists at all
scales and there is no quantum-classical (micro-macro) divide. This theory
verifiably solves all the foundational issues of QM in one stroke, while
affirming that the irreducible stochastic aspect is universal, consistent with
the validity of the principle of stationary action at all scales and
situations of dynamics \cite{Unni-RQM}. In a world in which the action
principle is operative, the action-waves and the fundamental uncertainty
$\Delta S\geq\hbar$ are inevitable; then, \emph{there cannot be any
dispersion-free ensemble, even in the macroscopic world}.

It is also clear now that misinterpreting the Schr\"{o}dinger equation for the
evolution of probability density as the dynamical equation for single
particles is the reason for the unphysical features of Bohm's theory as well.
In effect, the probability density of the whole statistical ensemble was made
to affect the dynamics of the single particle, through the fictitious and
nonlocal quantum potential (without a source).

\section{Summary}

In this paper I discussed several results that uphold the well-known assertion
made by J. von Neumann on the impossibility of hidden variable descriptions of
quantum mechanics, even at the level of single particle quantum mechanics. The
central result is the proof that Neumann's assumption of the linear additivity
of the expectation values, $Exp(aR+bS)=aExp(R)+bExp(S)$, is indeed a relation
of general validity, obeyed by all physical quantities. This resurrects
Neumann's famously contested proof of the impossibility of any hidden variable
description of quantum mechanics. The much discussed criticisms of this
assumption, by G. Hermann and J. S. Bell, were not valid. The second result is
the failure of Bell's counter-example to Neumann's theorem, a hidden variable
model of the measurements of the spin projection of a spin-1/2 particle. This
was demonstrated in different ways. Another related result discussed is that
the local hidden variable theories, for which Bell derived the inequalities,
are unphysical theories because they are not compatible with the fundamental
conservation laws. The discussion of the correlation functions of such
theories and the subsequent experimental tests of these intrinsically
unphysical theories were prompted and supported by Bell's critique of
Neumann's proof and his counter-example, both of which were incorrect. At this
point, it was necessary to present a clarificatory proof that the `causality
incompleteness' that motivated the work on hidden variable theories is very
different from the `EPR-incompleteness' of QM, discussed by Einstein, Podolsky
and Rosen in 1935. Finally, I identified the core reason for the impossibility
of dispersion-free ensembles in the relation between the quantum dispersion
and the action function. This result pre-empts any misconception that a
deterministic description of mechanics will ever be possible at any scale,
except as an approximation.

\section*{Acknowledgements}
This paper is the seed for a more complete work on the development of the conceptual core of quantum mechanics, bringing out the exact reasons for its present paradoxical state of immense success and withered foundations. The invaluable help from Martine Armand in the translations and language editing has been a significant factor in the general readability of the contents for the non-expert.

\end{document}